\documentclass[aps,showpacs,prl,preprint]{revtex4-1}
\usepackage{epsfig}
\usepackage{amsmath}
\usepackage{hyperref}

\begin{document}
\title{Phase diagram for nanostructuring CaF$_2$ surfaces by slow highly charged ions}
\author{A.S. El-Said}
\thanks{e-mail: a.s.el-said@hzdr.de}
\affiliation{Institute of Ion Beam Physics and Materials Research, Helmholtz-Zentrum Dresden-Rossendorf, 01328 Dresden, Germany, EU \\ Physics Department, Faculty of Science, Mansoura University, 35516 Mansoura, Egypt}
\author{R.A. Wilhelm}
\author{R. Heller}
\author{S. Facsko}
\thanks{e-mail: s.facsko@hzdr.de}
\affiliation{Institute of Ion Beam Physics and Materials Research, Helmholtz-Zentrum Dresden-Rossendorf, 01328 Dresden, Germany, EU}
\author{C. Lemell}
\author{G. Wachter}
\author{J. Burgd\"orfer}
\affiliation{Institute for Theoretical Physics, Vienna University of Technology, Wiedner Hauptstr.\ 8-10, A-1040 Vienna, Austria, EU}
\author{R. Ritter}
\author{F. Aumayr}
\thanks{e-mail: aumayr@iap.tuwien.ac.at}
\affiliation{Institute of Applied Physics, TU Wien - Vienna University of Technology, 1040 Vienna, Austria, EU}

\date{\today}

\begin{abstract}
Impacts of individual slow highly charged ions on alkaline earth halide and alkali halide surfaces create nano-scale surface modifications. For different materials and impact energies a wide variety of topographic alterations have been observed, ranging from regularly shaped pits to nano-hillocks. We present experimental evidence for a second threshold for defect creation supported by simulations involving the initial electronic heating and subsequent molecular dynamics. From our findings a unifying phase diagram underlying these diverse observations can be derived. By chemically etching of CaF$_2$ samples after irradiation with slow highly charged ions both above and below the potential energy threshold for hillock formation another threshold exists above which triangular pits are observed after etching. This threshold depends on both the potential and kinetic energies of the incident ion. Simulations indicate that this second threshold is associated with the formation of defect aggregates in the topmost layers of CaF$_2$.
\end{abstract}
\pacs{79.20.Rf,34.35.+a,61.80.Jh}
\maketitle

Studies of the interactions of slow ($v < v_\mathrm{Bohr}$), highly charged ions (HCI) with solid surfaces were originally aimed at gaining an understanding of the dynamical processes governing neutralization, relaxation and eventual dissipation of the very high potential energy density ($\sim$ keV/\AA$^3$) within a few femtoseconds \cite{bria90,meye91,burg91,sche99,elsa08,okab11}. This potential energy carried into the collision is given by the sum of all binding energies of electrons missing. More recently, the focus has shifted to material science driven applications, specifically to the development of novel techniques for material modification \cite{hamz01,tona07_1,pome07,saku07,lake11} and improved surface analysis \cite{hamz99,tona05,taka05,tona10}. Various types of surface nanostructures such as nano-sized hillocks, pits or craters have so far been observed after impact of individual HCI on different materials \cite{elsa08,tona07_2,tona08,hell08}. Their topography, appearance, and stability seem to depend sensitively on the material properties as well as on the potential energy, charge state, and kinetic energy of the incident ion (for a recent review see \cite{auma11}).

Surprisingly, even for very similar prototypical wide-bandgap insulators, ionic crystals of alkali halides and alkaline earth halides, vastly different and seemingly contradictory results have been found. Irradiation of KBr single crystals by individual highly charged Xe ions leads to formation of pits of one atomic layer depth \cite{hell08} while irradiation of CaF$_2$ single crystals produces nanometer high hillocks protruding from the surface \cite{elsa08}. In both cases it could be demonstrated that the surface nanostructures are the result of individual ion impacts, i.e.\ every structure is caused by the impact of a single ion only. Additionally, in both cases a threshold potential energy of the projectile had to be surpassed before the nanostructure could be observed. However, while for KBr this threshold potential energy for pit formation strongly decreases with increasing kinetic energy of the HCI \cite{hell08}, for hillock formation in CaF$_2$ it only slightly yet noticeably increases with increasing kinetic energy \cite{elsa08}. 

In this letter we present experimental evidence and results of simulations which supply the missing pieces to this puzzle and allow us to construct a phase-diagram as a function of kinetic and potential energies for the formation of different nanostructures. Key is the search for previously unobserved ``hidden'' surface structures after irradiation by ions with potential energies below the threshold for nano-hillock formation. By etching the samples we discover a second threshold at lower potential energy above which CaF$_2$ undergoes a nano-scale structural transformation even though not evident as a topographic change. It becomes, however, visible in the form of triangular pits after chemical etching. This threshold depends on both the potential energy and the kinetic energy of the HCI closely resembling the threshold behavior found for pit formation on KBr. Accompanying molecular dynamics simulations suggest this second threshold to be associated with lattice defect aggregation in CaF$_2$ following electronic excitations caused by the HCI-surface interaction.

Thin platelets of CaF$_2$ were prepared by cleaving a high purity single-crystal block grown from melt in an inert atmosphere along the (111) plane. This cleavage is known to produce atomically flat fluorine-terminated surfaces which are ideal for observing surface topographic changes down to the nanometer scale \cite{motz09}. $^{129}\mathrm{Xe}^{q+}$ ions were extracted from the electron beam ion trap (EBIT) at the Two-Source-Facility of the Helmholtz-Zentrum Dresden-Rossendorf using an electrostatic potential of 4.5\,kV. By using a two stage-deceleration system and adjusting the potential difference between source and target from 4.5\,kV down to 0.18\,kV, highly charged Xe$^{q+}$ projectiles over a wide range of charge states ($10 \leq q \leq 33$, corresponding to potential energies of 0.8\,keV $\leq E_\mathrm{pot} \leq  21.2$\,keV) and kinetic impact energies (6\,keV $\leq E_\mathrm{kin} \leq  150$\,keV) could be produced. The applied ion fluences were chosen between 0.5 and $5 \times 10^8$\,ions/cm$^2$, small enough to avoid overlapping of impact sites and high enough to obtain reasonable statistics. The time-averaged beam flux varied between $10^4 - 10^5$\,ions/s. The surfaces of the irradiated samples were investigated using atomic force microscopy (AFM) (Veeco Multimode). The AFM was operated in contact mode with a constant loading force of less than 5\,nN using non-conductive Si$_3$N$_4$ sensors (Veeco Instruments) with cantilevers of force constants $\sim 0.1$\,N/m. The image processing was performed using the WSxM software \cite{horc07}. Ion-irradiated CaF$_2$ samples were chemically etched using a HNO$_3$ solution (10\%\,vol.) at room temperature without agitation \cite{motz09}. Each platelet was immersed once in the etchant, subsequently into de-ionized water, and was finally dried in a stream of dry nitrogen. We have used much shorter etching times $t_\mathrm{e}$ than applied in standard etching techniques. For the latter, typically $t_\mathrm{e}\gtrsim 1$\,minute yields etch pits even starting from randomly occurring atomic-scale dislocations. Due to the dramatically enhanced etching speed in regions with a high defect density caused by the HCI, $t_\mathrm{e}=10$\,s turned out to be the optimum etching time combining good visibility of etch pits in AFM while selecting only defect clusters created by HCI impact. 

\begin{figure}
\centerline{\epsfig{file=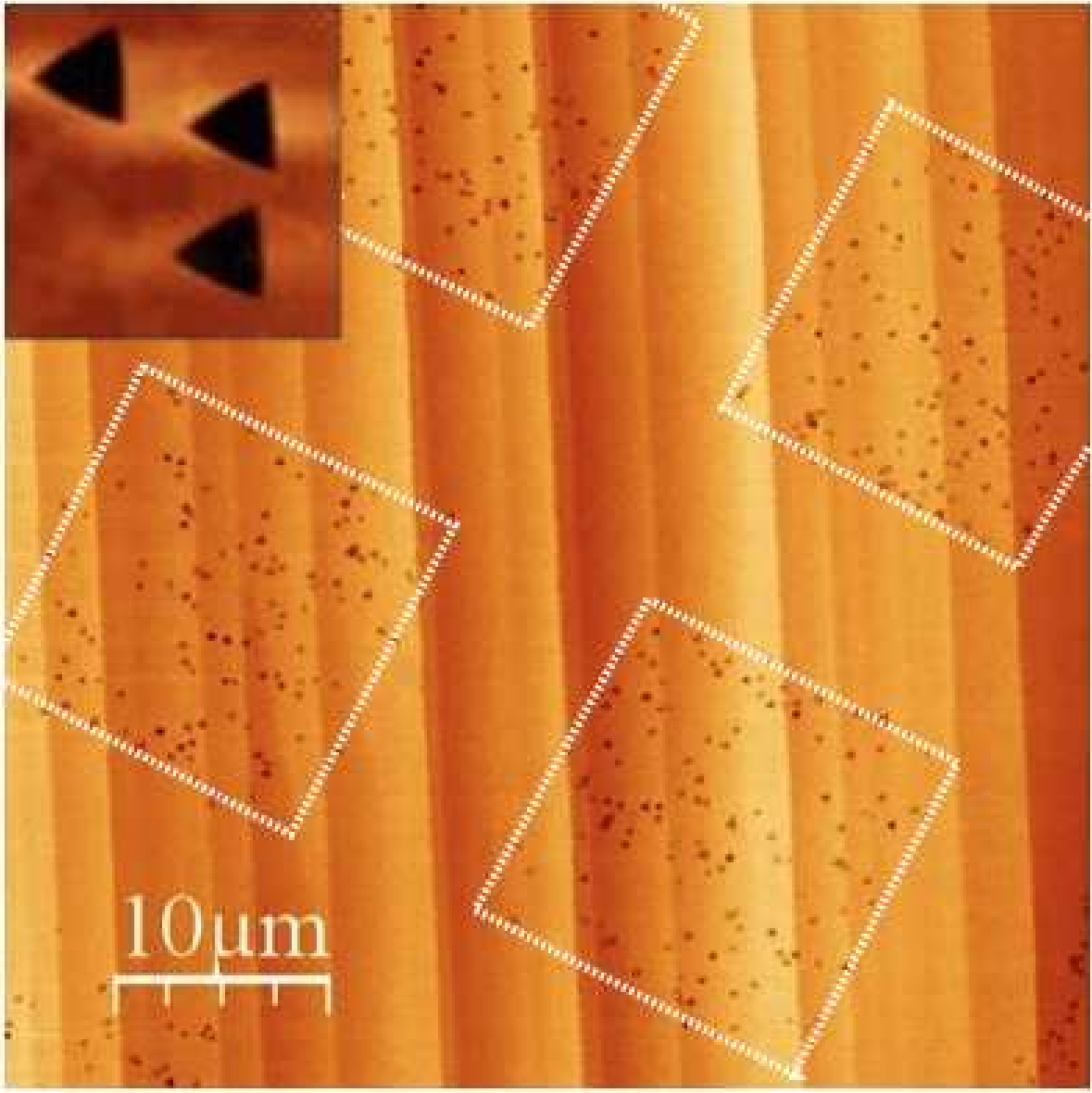,width=\columnwidth}}
\caption{(Color online) AFM topographic image ($50\times 50\, \mu\mathrm{m}^2$) of a CaF$_2$ surface showing etch pits after exposure to 150 keV Xe$^{33+}$ ions. The sample was irradiated through a mask (indicated by dotted lines) and subsequently chemically etched using HNO$_3$. The inset in the upper left corner shows a magnification of the etch pits ($1.5\times 1.5\, \mu\mathrm{m}^2$).}
\label{fig1}
\end{figure}
The observation of a pattern of well-defined irradiated and masked areas (Fig.\ \ref{fig1}) for 150 keV Xe$^{33+}$ ion impact on CaF$_2$ (111) is a direct evidence of HCI induced surface defects which can be clearly distinguished from randomly occurring dislocations and surface damage. In irradiated areas, etch pits of regularly structured 3-faced symmetric triangular depressions appear, which are similar to those observed after irradiation and etching of BaF$_2$ \cite{elsa10}. This particular geometrical shape originates from the (111) crystal lattice orientation of the CaF$_2$ sample \cite{motz09}. The number of pits is in good agreement with the applied ion fluence, i.e.\ each etch pit is created by a single ion impact. We suppose the pits are localized at the sites where HCI impact created hillocks were situated prior to etching. The charge state ($q=33$) of the incident ion corresponds to a potential energy well above the threshold for nano-hillock formation.

\begin{figure}
\centerline{\epsfig{file=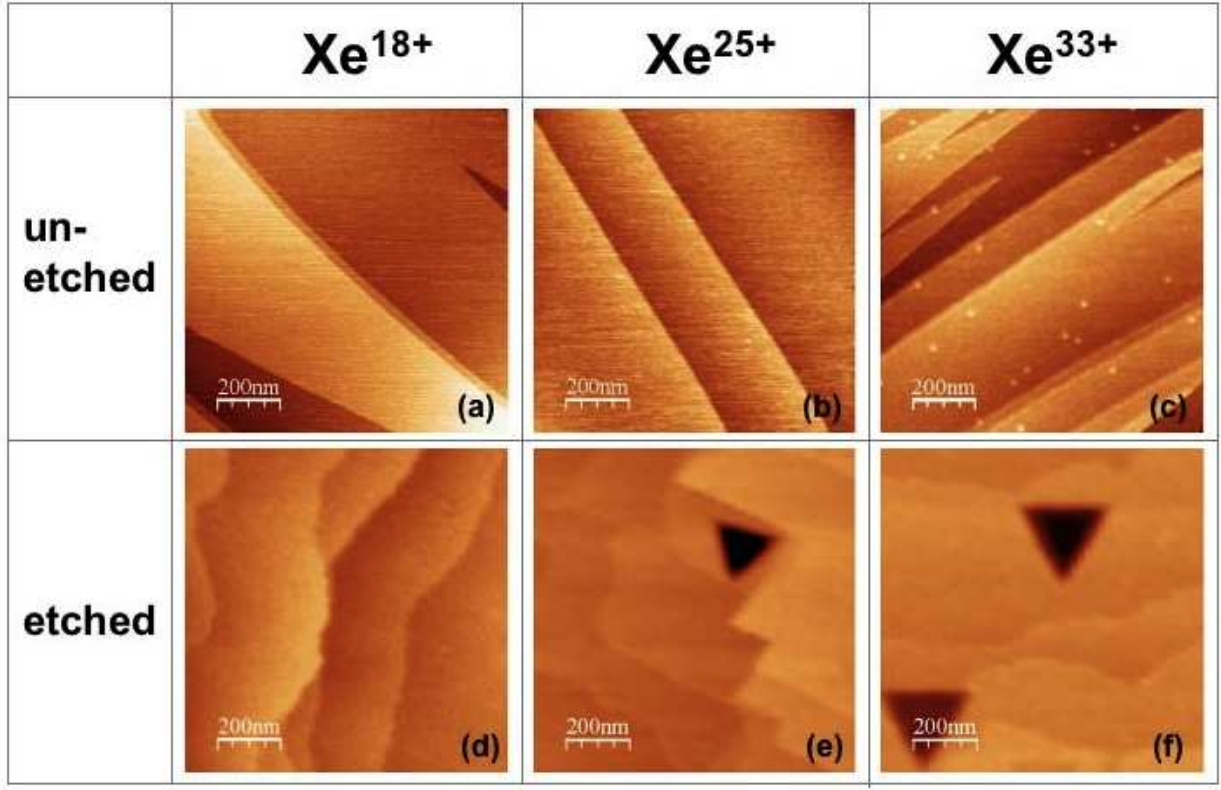,width=\columnwidth}}
\caption{(Color online) Topographic contact-mode AFM images of CaF$_2$ (111) samples irradiated by 40 keV Xe ions in different charge states (columns): (a, d) Xe$^{18+}$, (b, e) Xe$^{25+}$, and (c, f) Xe$^{33+}$. In each frame an area of 1 $\mu$m $\times 1\,\mu$m is displayed. Upper row: resulting images without etching (a, b, c), lower row: images after etching by HNO$_3$ (d, e, f). Ion fluences were $2 \times 10^8$ ions/cm$^2$ for (e, f) and $1 - 2 \times 10^9$ ions/cm$^2$  for (a, b, c, d).}
\label{fig2}
\end{figure}

Lowering the charge state to values below the potential energy threshold for hillock formation ($q_\mathrm{th}\approx 28$ for Xe; $E_\mathrm{pot}=12$ keV) reveals the appearance of similar pits in the absence of preceding hillocks (Fig.\ \ref{fig2}).
At the same kinetic energy of $E_\mathrm{kin}=40$\,keV for ``low'' charge states ($q\lesssim 18$) no damage of the etched surface is visible, whereas at higher charge state ($q=25,\, E_\mathrm{pot}=8.1$\,keV) etch pits appear.

\begin{figure}
\centerline{\epsfig{file=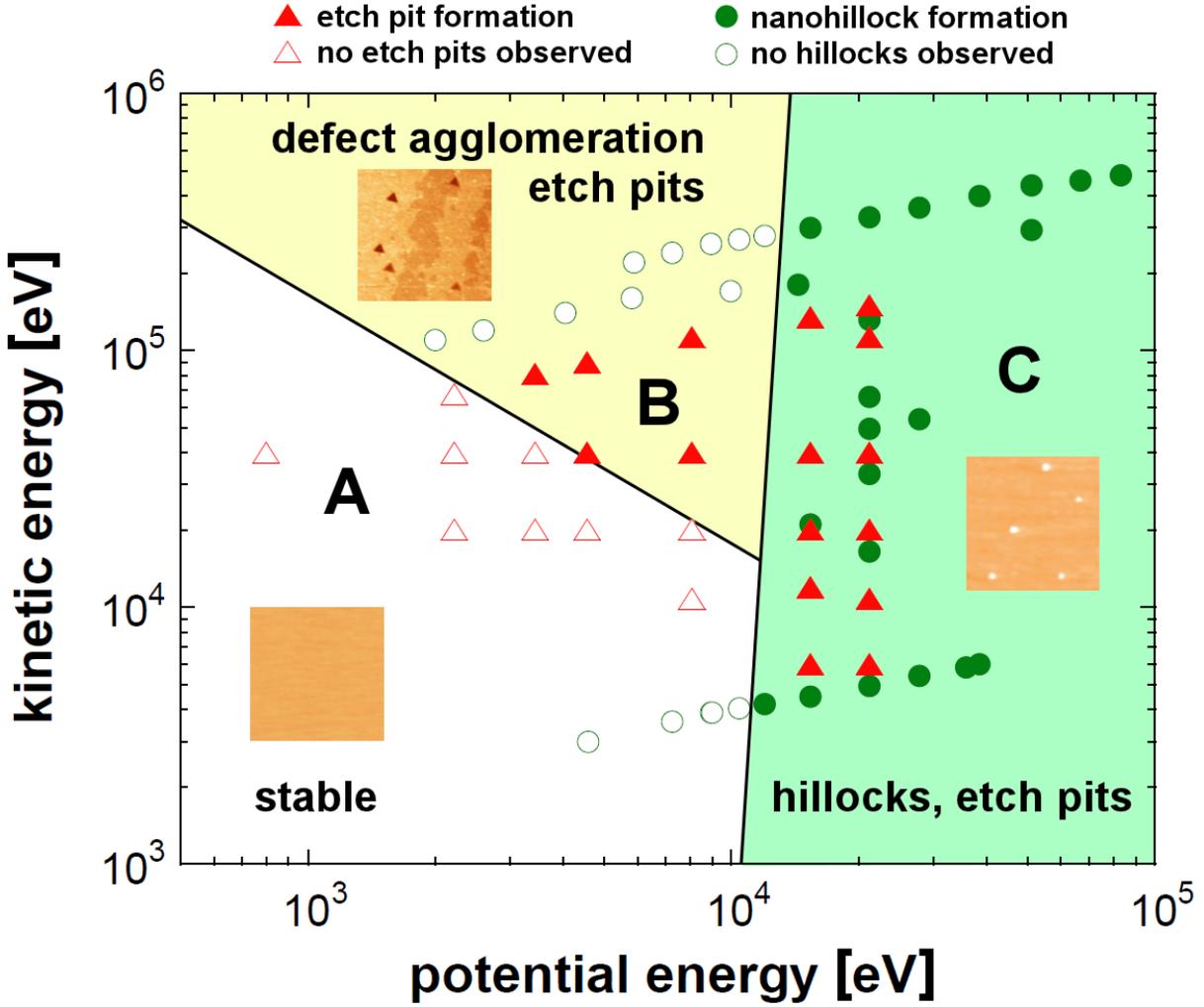,width=\columnwidth}}
\caption{(Color online) Hillock and etch pit formation on CaF$_2$ (111) induced by irradiation with highly charged Xe ions. Full (open) circles show pairs of potential and kinetic energies where hillocks are produced (absent) after irradiation, full (open) triangles indicate pairs where pits are present (missing) after etching the irradiated samples.}
\label{fig3}
\end{figure}

In order to investigate the influence of both potential and kinetic energies on etch pit formation, we performed systematic irradiations with $^{129}\mathrm{Xe}^{q+}$ projectiles of different charge state ($q = 10$ to 33) and with varying kinetic energy on CaF$_2$. The resulting surface damage and modification can be summarized by a phase-diagram with potential and kinetic energies as state variables (Fig.\ \ref{fig3}). 
Three different ``phases'' pertaining to surface structuring can be distinguished: the stability region (area A) within which no significant surface modification detectable by AFM can be induced by HCI impact (Figs.\ \ref{fig2}a, d), the defect agglomeration region B which becomes visible as regularly shaped pits for CaF$_2$ only upon etching (Figs.\ \ref{fig2}b, e), and the nano-hillock phase C (Figs.\ \ref{fig2}c, f) in which hillocks result from nano-melting. Etch pits become visible in Phase C after etching and we note that we did not find a single case for which the hillock was not accompanied by an etch pit.

It is well known that the threshold for hillock formation strongly depends on potential energy but only weakly on kinetic energy \cite{elsa08,auma11}, implying an almost vertical boundary of region C in Fig.\ \ref{fig3}. On the contrary, the border separating the stability regions A and the defect agglomeration region B (etch pits) is strongly dependent on both kinetic and potential energies. Ions with lower kinetic energy require more potential energy to create etch-able damage than faster ones. Such synergistic effects of kinetic and potential energies have been, indeed, observed for pit formation in KBr \cite{hell08}, however with the remarkable difference that no chemical etching was a prerequisite for the pits to be observed.

\begin{figure}
\centerline{\epsfig{file=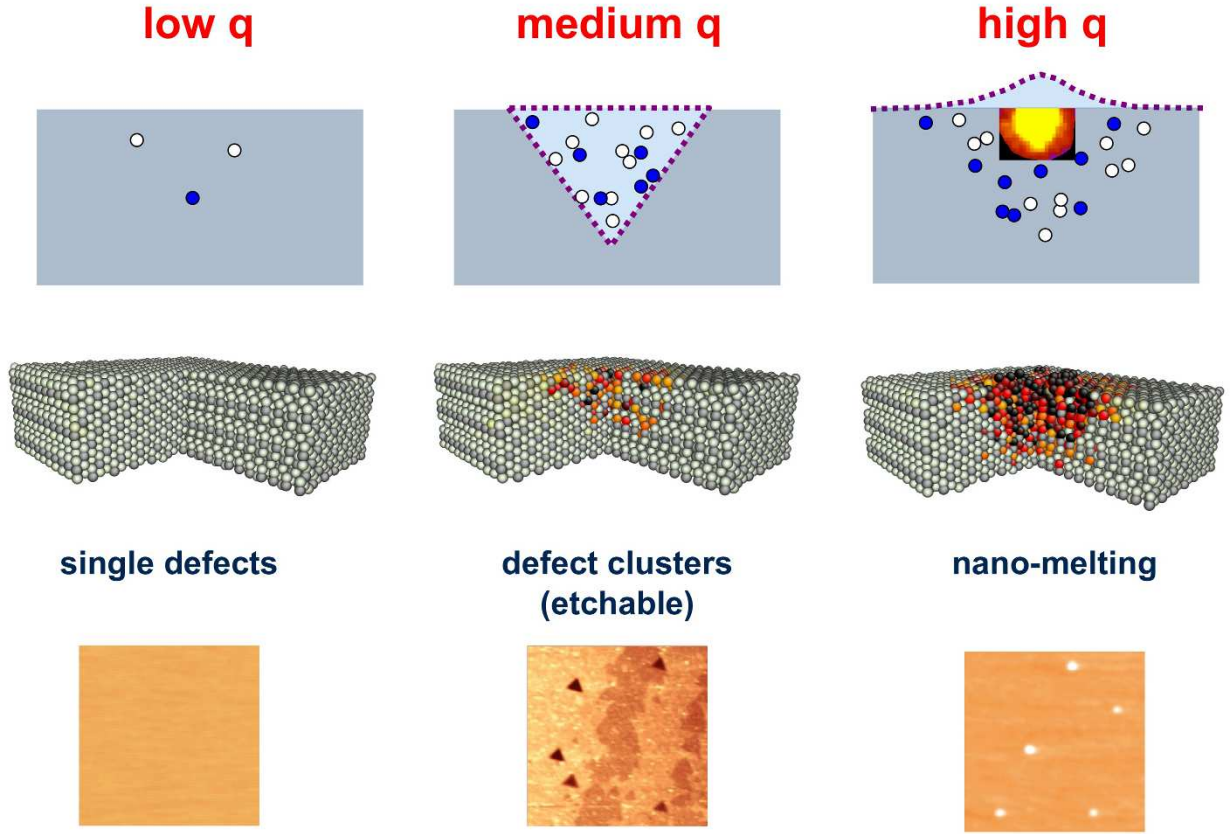,width=\columnwidth}}
\caption{(Color online) Scenario for surface modification as a function of charge $q$ or, equivalently, potential energy of the HCI. Upper row: the charge state controls the created surface modification from non-etchable single defects (low $q$) to defect aggregates (medium $q$) and to locally molten zones (high $q$). Lower row: AFM images. Center row: typical results of molecular dynamics simulations show that the initial electronic excitation of the surface and energy transfer to the lattice leads to a considerable number of displacements (center column) even before melting of the surface sets in (right column).}
\label{fig4}
\end{figure}

These experimental findings suggest the following scenario for nanostructure formation on alkaline earth halides and alkali halides qualitatively supported by simulations involving a sequence of three steps: initial heating of electrons by multiple electron transfer and Auger relaxation, hot electron transport and dissipation with accompanying lattice heating by electron-optical phonon coupling, and subsequent molecular dynamics (for details see \cite{wach09}). It should be noted that for ionic crystals, in particular far from charge equilibrium, accurate binary potentials are not available and quantitatively reliable predictions are not feasible. Nevertheless, the following qualitative trends can be readily extracted. For HCI in ``low'' charge states (Fig.\ \ref{fig4}, left) only a few (i.e.\ low density) individual defects (point defects, single vacancies) are created at or below the surface.
These defects either remain below the surface, easily anneal or are too small to be detected by means of AFM. Since the etchability of CaF$_2$ is strongly coupled to the creation of large defects aggregates \cite{trau98} rather than to point defects, no pits are observed after etching. Our MD simulations do not yield any significant number of lattice displacements for low $q$.

For larger $q$ and, correspondingly, larger energy, the potential sputtering yield strongly increases \cite{auma04} as well as the density of defects (excitons, color centers), which is now large enough leading to defect clusters and aggregates (Fig.\ \ref{fig4} center column). Depending on their mobility, defects may diffuse to the surface, lead to defect mediated desorption \cite{hayd99} and thus form (monatomic) pits as observed in the case of the alkali halide KBr \cite{hell08}. The defect mediated desorption mechanism is less probable in CaF$_2$, since color center recombination below the surface is much more likely \cite{will89} due to the small energy gain of color center pair formation as well as the formation of more complex (and therefore immobile) defect agglomerates \cite{rix09, rix11}. The material in the vicinity of the impact region is not ablated but structurally weakened and forms the nucleus of an etchable defect subsequently removed by a suitable etchant \cite{elsa10}. The synergistic effect induced by the accompanying kinetic energy originates from kinetically induced defects created in the collision cascade which enhance the trapping of the color centers created by potential energy \cite{hayd01} and therefore increases defect agglomeration. Consequently, the borderline between the regions A (stable) and B (etchable surface defects) has a negative slope in the phase diagram (Fig.\ \ref{fig3}). While our MD simulation cannot directly account for the defect cluster formation (due to the lack of realistic binary potentials for color centers and charge-exchanged constituents), it does predict a large number of atomic displacements (Fig.\ \ref{fig4} center) believed to be a necessary precursor for defect aggregation.

At still higher potential energies (Fig.\ \ref{fig4} right column), heating of the lattice atoms by primary and secondary electrons from the deexcitation of the HCI surpasses the melting threshold of the solid \cite{elsa08,wach09}. Heat and pressure deforms the surface and after rapid quenching a hillock remains at the surface. With increasing kinetic energy, the region where the potential energy of the HCI is deposited, extends slightly deeper into the target \cite{elsa08}. Therefore, the kinetic energy dependence of the borderline between the region of nano-hillock formation (region C) and defect clustering without protrusion (region B) is only weak with a slightly positive slope. The overall surface damage (lattice distortion, defect aggregations) extends well beyond the molten core. While the latter determines the diameter of the hillock, the former determines the size of the nucleus of the etch pit.

Even though the present scenario is demonstrated specifically for CaF$_2$, we surmise that it should hold for other halide crystals as well. While borderlines between different regions A, B, and C will, of course, depend on the specific target material, we expect the phase diagram (Fig.\ \ref{fig3}) to remain qualitatively valid. For BaF$_2$ (111) and KBr (001) for example, we have previously observed only the A and B phases \cite{elsa10,facs09}. The phase diagram predicts that by further increasing the potential energy of the HCI we should be able to reach region C, i.e.\ hillock formation (or melting). Indeed, we very recently found first indications for hillock formation on BaF$_2$ \cite{elsa11}.

In summary, we have established a phase diagram for nano-scale surface modification of alkaline earth halides and alkali halides by highly-charged ion impact with its potential and kinetic energies as control parameters. In addition to the region of predominantly potential energy driven melting and hillock formation a second region was identified in which a sufficient number of defects agglomerate such that chemical etchants are able to remove material leaving triangular shaped pits on the surface. The etchability of the defect cluster not only depends on the potential energy of the HCI but also strongly on the kinetic energy of the projectile. This scenario seems to be generally applicable to other alkaline earth and alkali halide surfaces as well.

A.S.E. thanks Alexander von Humboldt Foundation for financial support. The Deutsche Forschungsgemeinschaft (DFG) is gratefully acknowledged for financial support under Proj.\ No.\ HE 6174/1-1. R. R. is a recipient of a DOC-fellowship of the Austrian Academy of Sciences. Support from the Austrian Science Foundation FWF under Proj.\ No.\ SFB-041 ViCoM  and the International Max Planck Research School of Advanced Photon Science APS (G.W.) is acknowledged.

\end{document}